\documentclass[prb,aps]{revtex4}
\usepackage{graphicx}
\usepackage{comment}
\usepackage{bm}
\usepackage{amsmath,amssymb} 
\usepackage{epstopdf}
\usepackage{verbatim}
\usepackage{float}
\newcommand{\gp}{G_{\infty,{\rm p}}}
\newcommand{\kp}{K_{\infty,{\rm p}}}
\newcommand{\mop}{M_{\infty,{\rm p}}}
\newcommand{\ga}{G_{\infty,{\rm af}}}
\newcommand{\ka}{K_{\infty,{\rm af}}}

\begin{document}

\title{The instantaneous shear modulus in the shoving model}
\author{Jeppe C. Dyre}
\email{dyre@ruc.dk}
\affiliation{DNRF Centre ``Glass and Time'', IMFUFA, Department of Sciences, Roskilde University, Postbox 260, DK-4000 Roskilde, Denmark}
\author{Wei Hua Wang}
\email{whw@iphy.ac.cn}
\affiliation{Institute of Physics, Chinese Academy of Sciences, Beijing 100190, P.O. Box 603, China}

\date{\today}

\begin{abstract}
We point out that the instantaneous shear modulus $G_\infty$ of the shoving model for the non-Arrhenius temperature dependence of viscous liquids' relaxation time is the experimentally accessible high-frequency plateau modulus, not the idealized instantaneous affine shear modulus that cannot be measured. Data for a large selection of metallic glasses are compared to three different versions of the shoving model. The original shear-modulus based  version shows a slight correlation to the Poisson ratio, which is eliminated by the energy-landscape formulation of the model in which the bulk modulus plays a minor role.
\end{abstract}
\maketitle

\section{Introduction}

In a recent Communication\cite{puo12}Puosi and Leporini showed from computer simulations that the relevant high-frequency shear modulus controlling the relaxation is not the idealized shear modulus corresponding to affine deformations at truly infinite frequency. Rather, it is the shear modulus referring to time scales that are on the one hand much shorter than any relaxation time, but on the other hand much longer than typical vibration times. This confirms findings by other groups.\cite{sta02,rib09,rib11} Puosi and Leporini further proposed an extension of the shoving model to allow for heterogeneities and showed that the new model fits simulation data very well. This extension is consistent with previous works by Khronik {\it et al.}, who introduced the idea of a distribution of local shear moduli to explain sub $T_g$ relaxations within the shoving-model framework.\cite{kho09,mit11}

Commenting on the instantaneous shear modulus of the shoving model Puosi and Leporini wrote: ``It is quite apparent that $G_\infty$, the central quantity of the standard elastic models, poorly correlates with the structural relaxation time.'' This conclusion derives from the understanding that the shoving model and related elastic models are based on the idealized affine infinite-frequency shear modulus, not the experimentally measurable high-frequency plateau modulus, which was traditionally used for comparing model to experiment. Unfortunately, theorists and experimentalists have used for many years the same symbol, $G_\infty$, for different physical properties. In experiment the term ``instantaneous shear modulus''  always meant the shear modulus at the highest obtainable frequencies (typically MHz or GHz, or even high kHz, depending on the technique used),\cite{bar67,har76,nos96,iiz05,rod07,sle08,pau10} where it generally becomes frequency independent and is usually -- though not always -- denoted by $G_\infty$ .\cite{bar67,har76} In his excellent text book {\it The dynamic properties of supercooled liquids} from 1976 Harrison refers to this quantity as ``the high-frequency limiting shear modulus''.\cite{har76} Likewise, whenever the Maxwell relaxation time $\tau_M=\eta/G_\infty$ in the literature has been calculated for a liquid, $G_\infty$ was always identified as the experimental limiting high-frequency shear modulus of the plateau, much below phonon frequencies.

In liquid-state theory, $G_\infty$ was traditionally the idealized, truly infinite-frequency limit of the fluctuation-dissipation theorem expression,\cite{zwa65,han06} proportional to the mean-square shear-stress equilibrium fluctuation. Previously, there was no reason to believe that these two quantities should differ in any significant way. It now turns out that for some systems, as temperature changes at constant density, one quantity increases and the other decreases.\cite{puo12,yos10} The purpose of the present paper is to show that the shoving model derivation assumes $G_\infty$ is the quantity, which Puosi and Leporini referred to as the plateau modulus.\cite{puo12,sta02,rib09,rib11}  We suggest a consistent notation for the two instantaneous shear moduli, which can be used whenever there is a risk of confusing them.\cite{puo12,yan11} Finally, we reanalyze data for a large selection of metallic glasses in order to compare to predictions of three versions of the shoving model. 

Section II reviews the shoving model, Sec. III recalls the model energy-barrier calculation and relates it to the reversible-work theorem of statistical mechanics. Section IV compares model predictions to data, and Sec. V gives a brief summary.

\section{The shoving model}

The shoving model for the non-Arrhenius temperature dependence of the main (alpha) relaxation time $\tau$ of a glass-forming liquid predicts that $\tau(T)=\tau_0\exp\left[G_\infty(T)V_c /(k_BT)\right]$.\cite{dyr96,dyr98a} Here $\tau_0$ is a prefactor of order 0.1 picosecond, $G_\infty$ is the ``instantaneous'' shear modulus, $V_c$ is a characteristic volume of order a molecular volume, $k_B$ is Boltzmann's constant, and $T$ is the temperature. In order to minimize the number of free parameters, the shoving model assumes {\it ad hoc} that $V_c$ is temperature independent. In this way the model connects directly two experimentally measurable quantities, $\tau(T)$ and $G_\infty(T)$. The model does not consider what causes the unusually large temperature dependence of glass-forming liquids' instantaneous shear moduli. This problem was addressed recently in interesting papers by Brito and Wyart, who proposed that the increase of the instantaneous shear modulus derives from a stiffening of the Boson peak as temperature is lowered.\cite{bri09,wya10}

The shoving model and related elastic models have been confirmed for a number of organic, oxide, chalcogenide, oxynitride, and metallic liquids,\cite{dyr96,dyr06b,mag08,tor09,rou11,wan11,xu11, wan12}, as well as in some computer simulations,\cite{puo12,sta02,rib09,rib11} but failures of the model have also been reported.\cite{gra98,buc09} Most experimental conformations relate to a liquid's temperature-dependent equilibrium relaxation time, but there are also tests confirming the shoving and related elastic models for aging experiments.\cite{dyr98,kho08,kho09,mit11} More data are certainly needed before it is clear whether the shoving and related models account generally for the non-Arrhenius temperature dependence of $\tau$ observed in supercooled liquids. For reviews of elastic models the reader is referred to Refs. \onlinecite{dyr06a} and \onlinecite{nem06}.

The shoving model is based on the assumption that the main contribution to the activation energy for a ``flow event'' -- a rearrangement from one potential energy minimum to another -- is the work done in shoving aside the surroundings in order to increase the volume available for rearranging the molecules. The model assumptions are:\cite{dyr96,dyr06b,dyr06a}

\begin{itemize}
\item The main contribution to the activation free energy is elastic energy.
\item This elastic energy is located in the surroundings of the rearranging molecules.
\item The elastic energy is shear elastic energy, i.e., not associated with density changes. 
\end{itemize} 
To make things simple, the shoving model assumes spherical symmetry of a flow event and calculates the activation free energy as the work done in shoving aside the surroundings (expanding a sphere) in order to create room for a flow event. 

Is it reasonable to assume that the main contribution to the activation energy comes from the surroundings? What about contributions from the rearranging molecules themselves? A simple argument shows that the former contribution dominates.\cite{dyr98} Suppose that rearranging at constant volume is energetically very costly because the molecules are forced into close contact during the rearrangement process; this is the main physical idea of the shoving model. In this case, allowing for just a slightly larger volume for the rearranging molecules implies a considerable lowering of the energy cost. If the radius change is $\Delta r$ and the energy barrier contribution from the rearranging molecules within the sphere is $f(\Delta r)$, the total energy barrier involves a further quadratic contribution from deforming elastically the surroundings: $\Delta E =f(\Delta r)+A(\Delta r)^2$. The fact that the function $f(\Delta r)$ decreases significantly when $\Delta r$ increases slightly above zero is expressed mathematically as $|d\ln f/d\ln\Delta r|\gg 1$. Optimizing $\Delta r$ in order to find the lowest barrier leads to $f'(\Delta r)+2A\Delta r=0$. Thus the ratio between the ``shoving'' work and the ``inner'' contribution is 
$A(\Delta r)^2/f(\Delta r)=-f'(\Delta r)/[2\Delta rf(\Delta r)]=|d\ln f/d\ln\Delta r|/2 \gg 1$.

It is the high-frequency {\it shear} elastic constant that enters into the shoving model prediction because the expansion of a sphere in an elastic solid results in a radial displacement in the surroundings, which varies with distance $r$ to the sphere center as $r^{-2}$.\cite{lan70} This is a pure shear deformation, i.e., with zero divergence and thus no density changes anywhere (compare to the Coulomb electric field of a point charge $\propto r^{-2}$, which also has zero divergence). Less idealized geometries would result in some density change and thus also involve the bulk modulus, but it has been shown generally that the bulk elastic energy of the far field of an elastic dipole constitutes a most 10\% of the total elastic energy.\cite{dyr07} This confirms the interesting phenomenon of ``shear dominance'' noted some time ago in various contexts of condensed matter physics and materials science (see Refs. \onlinecite{dyr07} and \onlinecite{gra92} and their references).

\section{Calculating the free energy barrier using the reversible-work theorem of statistical mechanics}

The idealized instantaneous affine shear modulus refers to the hypothetical situation where one imposes an instantaneous, perfectly affine shear deformation on the system, corresponding to a time scale so fast that the atoms do not move, i.e., on a femtosecond time scale. This quantity is not experimentally accessible for the following reasons. It is difficult to imagine imposing an affine shear deformation on a system on a femtosecond time scale. Even if this were possible, one would not observe a high-frequency limiting shear modulus above THz frequencies -- inertial effects would set in and cause the modulus to go to zero as frequency diverges without limit. On the other hand, in a computer simulation an instantaneous affine shear deformation is easily imposed on a system.

The energy barrier is calculated as the work done in shoving aside the surroundings. According to a fundamental theorem of statistical mechanics, the difference in free energy between two states can be calculated as the reversible isothermal work done to bring the system from one to the other state. Thus, in calculating the energy barrier, the expansion of the sphere must take place so slowly that there is equilibrium in the surroundings throughout the expansion process. On the other hand, since the activation (free) energy refers to the thermally activated creation of extra volume in the fixed, glassy structure of the surrounding molecules, the expansion must be fast enough that no relaxations take place in the surrounding liquid. This means that the time of expansion must be much smaller than the alpha relaxation time. In conclusion, the relevant elastic constants of elastic models refer to high frequencies, still much below phonon frequencies. In this frequency range the bulk and shear elastic constants are usually frequency independent for highly viscous liquids (ignoring possible secondary relaxations). 

Over the years we have occasionally met the misunderstanding that the ``shoving'' process is the work done during the actual barrier transition of the rearranging molecules, which takes place over a few picoseconds. This is not correct; the (free) energy barrier is a difference in free energy between two states -- the starting state and the transition state -- and as detailed above this quantity is calculated by reference to statistical mechanics and the reversible-work theorem.

Both the experimentally measurable high-frequency plateau shear modulus -- referring to frequencies much below phonon frequencies -- and the experimentally non-accessible truly instantaneous affine shear modulus have traditionally been denoted by $G_\infty$. This would not present a serious problem if the two moduli were more or less identical or, from the shoving model perspective, if they were proportional in their temperature variation throughout the phase diagram. It now appears that neither is the case.\cite{puo12,yos10} This calls for introducing an unambiguous notation. We suggest denoting the experimental ``instantaneous'' shear modulus by $\gp$ (p for plateau) and the idealized, affine truly instantaneous shear modulus by $\ga$ (``af'' for affine). In this notation, which need only to be used whenever there is a risk of confusion, the shoving model prediction is

\begin{equation}\label{shov}
\tau(T)\,=\,\tau_0\exp\left[\frac{\gp(T)V_c}{k_BT}\right]\,.
\end{equation}

Before proceeding we note the fluctuation-dissipation (FD) theorem expressions for the two instantaneous shear moduli. Recall that if one defines $S_{xy}\equiv\sum_i x_i F_{y,i}$ where $x_i$ is the x component of the position vector of particle $i$ and $F_{y,i}$ is the y component on the force on this particle, the FD theorem expression for the frequency-dependent shear viscosity $\eta(\omega)$ is (where $V$ is the system volume)

\begin{equation}\label{fd}
\eta(\omega)
\,=\,\frac{\int_0^\infty \langle S_{xy}(0)S_{xy}(t)\rangle e^{-i\omega t}dt}{k_BT\,V}\,.
\end{equation}
Since the frequency-dependent shear modulus is related to the viscosity by $G(\omega)=i\omega\eta(\omega)$, letting frequency go to infinity in Eq. (\ref{fd}) gives the well-known expression for the idealized affine infinite-frequency shear modulus $\ga$,\cite{han06}

\begin{equation}\label{gainf_fd}
\ga
\,=\,
\frac{\langle S_{xy}^2\rangle}{k_BT\,V}\,.
\end{equation}
As mentioned, this quantity cannot be measured in experiment. The quantity that can be measured, the plateau modulus $\gp$, is given by the analogous expression where $S_{xy}$ is averaged over a few molecular vibration periods (i.e., over some picoseconds):

\begin{equation}\label{gpinf_fd}
\gp
\,=\,
\frac{\langle \overline{S}_{xy}^2\rangle}{k_BT\,V}\,.
\end{equation}

We finally note that $\ga/(\rho T)$ and $\gp/(\rho T)$ are both isomorph invariants. The invariance of the first expression was shown in Ref. \onlinecite{gna09}, that of the latter follows by analogous arguments. This shows that the shoving model survives the ``isomorph filter'' according to which any generally applicable model for the non-Arrhenius temperature dependence of a supercooled liquid's relaxation time must give this as a function of an isomorph invariant.\cite{gna09}

\section{Comparing the energy-landscape version of the shoving model to metallic glass data}

\begin{figure}[width=5cm]%
\centering
  \includegraphics{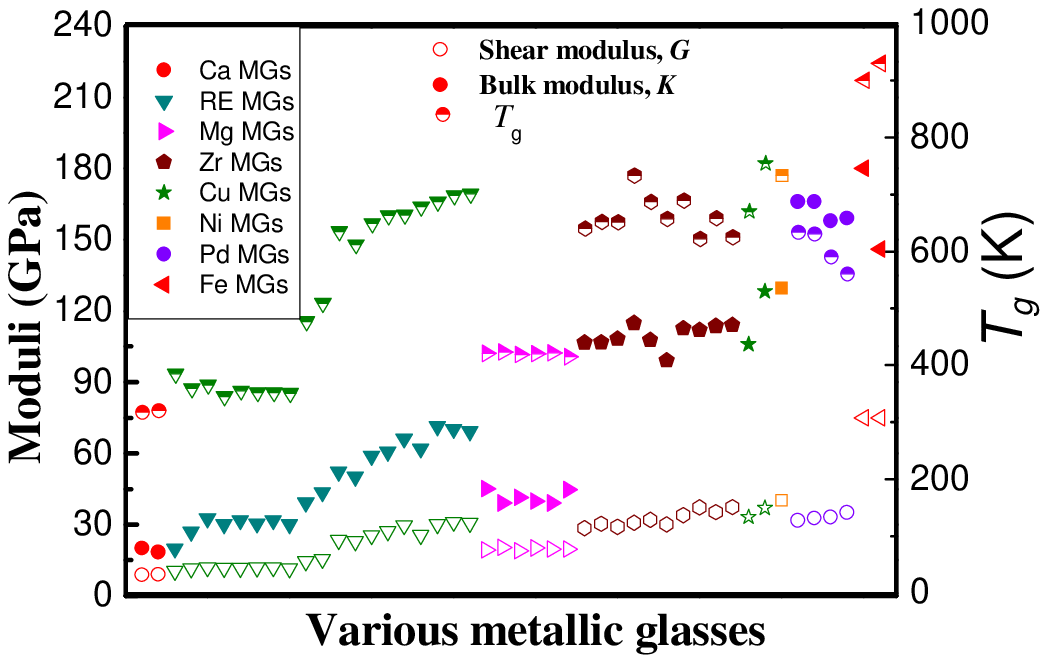}
   \caption{Shear and bulk moduli for a range of metallic glasses (open and full symbols, respectively, left axis) and the glass transition temperatures $T_g$ (half-full symbols, right axis). These data are used below for testing the different versions of the shoving model.\label{fig1}}
\end{figure}

The energy-landscape justification of the shoving model\cite{dyr04} is based on a classical argument, which estimates the barrier height for a jump between two (free) energy minima from the curvature at the minima, leading to $\ln\tau\propto 1/\langle u^2\rangle$ where $\langle u^2\rangle$ is the vibrational mean-square displacement.\cite{fly68,hal87,koh88,buc92} If the high-frequency shear and bulk plateau moduli, $\gp$ and $\kp$, differ from the ideal, affine moduli $\ga$ and $\ka$, the former are the relevant ones for the phonon spectra. In fact, one way to probe plateau moduli is to measure the linear (i.e., low-wavevector) parts of the phonon spectra or, equivalently, the shear and longitudinal high-frequency sound velocities. This is also how Ribero {\it et al.} probed $G_\infty$ in computer simulations.\cite{rib09,rib11} In this section term ``instantaneous'' moduli implies the plateau moduli.

In a simple, isotropic elastic model the vibrational mean-square displacement may be estimated from the instantaneous moduli by using the fact that for a given wavevector, there are two transverse and one longitudinal phonon. If $M=K+(4/3)G$ is the longitudinal modulus, this leads\cite{dyr04} to $\langle u^2\rangle/T\propto 1/\mop+2/\gp$. For the temperature dependence of the relaxation time this implies via $\ln\tau\propto 1/\langle u^2\rangle$

\begin{equation}\label{elp}
\tau(T)\,=\,\tau_0\exp\left[\frac{V_c'}{k_BT}\left(\frac{1}{\mop(T)}+\frac{2}{\gp(T)}\right)^{-1}\right]\,,
\end{equation}
where $V_c'$ is a molecular-sized volume. As shown in Ref. \onlinecite{dyr04}, this expression implies shear dominance for the temperature dependence of $\tau$. This was shown by first defining for any quantity $Q$ that increases as $T$ decreases  its temperature index as $I_Q\equiv -d\ln Q / d\ln T$. Equation (\ref{elp}) implies that the temperature index of $\tau$'s activation energy can be written $(1-\alpha)I_{\gp} +\alpha I_{\kp}$, where $0<\alpha < 0.08$ is obeyed no matter what is the ratio $\gp/\kp$.\cite{dyr04} In other words, at least 92\% of the non-Arrhenius temperature dependence of the relaxation time derives from $\gp$'s temperature dependence. The instantaneous bulk modulus $\kp$ plays only a minor role for three reasons: 

\begin{itemize}
\item Two phonons are transversal for each one that is longitudinal.
\item The instantaneous shear modulus $\gp$ affects also the longitudinal phonons.
\item Longitudinal phonons are harder than transverse and therefore contribute less than one third to the vibrational mean-square displacement.
\end{itemize} 
It follows from Eq. (\ref{elp}) that the temperature-dependent activation (free) energy $\Delta E(T)$ is given by

\begin{equation}\label{de}
\Delta E(T)
\,\propto\,\gp(T)V_c'\frac{\kp(T)+4\gp(T)/3}{2\kp(T)+11\gp(T)/3}\,,
\end{equation}

Traditionally the shoving model is compared to experiment by plotting the logarithm of the relaxation time (or, equivalently, the viscosity) as a function of $\gp(T)/T$.\cite{dyr96,dyr98a,dyr06b,mag08,tor09,rou11,buc09} Such a plot checks whether the predicted linear relationship between $\ln\tau$ and $\gp(T)/T$ is observed with a physically reasonable prefactor. A convenient way of making this plot is via a generalized Angell plot in which the x coordinate is normalized to unity at the glass transition by defining $X\equiv [\gp(T)T_g]/[\gp(T_g)T]$. Unfortunately, measuring the high-frequency plateau shear modulus of the equilibrium metastable supercooled liquid is difficult and not too many data are available even today. An alternative way of testing the shoving model is by making use of the facts that 1) for a fixed  cooling rate the relaxation time has a certain value at $T_g$, 2) the (DC) shear modulus of the glass, $G$, is almost temperature independent and equal to the instantaneous shear modulus of the liquid at $T_g$: $G\cong \gp(T_g)$. The glass shear modulus is easy to measure, and the prediction of the model is that $GV_c/k_BT_g\cong {\rm Const.}$, in which it is reasonable to assume that the constant is more or the same for similar systems. For the range of metallic glasses compared below to model predictions we further assume that the characteristic volume $V_c$ is proportional to the molar volume $V_m$ with a universal proportionality constant. The point of these simplifying assumptions is to eliminate all non-trivial free parameters.

Figure 1 gives bulk and shear modulus data for a range of metallic glasses, as well as their glass transition temperatures ranging from 317K for some of the Ca-based glasses to 930K for some Fe-based ones. Most of these data were discussed and compared to elastic model predictions in Refs. \onlinecite{wan11,wan12}, but the cupper-based glasses were replaced.

Figure 2 compares data to different versions of the shoving model [Eq. (\ref{shov})]. Since the glass transition takes place when the relaxation time upon cooling reaches a certain value, the standard shoving model prediction is 

\begin{equation}\label{shv1}
T_g\propto 
\Delta E(T_g)\propto G\,V_m \,
\end{equation}
with a universal proportionality constant. Figure 2(a) tests this prediction based on the data of Fig. 1, where the metallic glasses are sorted according to their Poisson ratio ($R$ is the gas constant). Given the diversity of the glasses, the model's simplicity, and the simplifying assumption that $V_c\propto V_m$ with a universal proportionality constant, the data show good agreement with the shoving model. There is, as noted in Refs. \onlinecite{wan11} and \onlinecite{wan12}, some correlation with Poisson's ratio.\cite{gre11} Before addressing whether the energy-landscape version of the shoving model rectifies this, we show in Fig. 2(b) how data compares to the elastic model in which $G$ is replaced by $K$. Here the correlation with Poisson's ratio is much stronger. Altogether, Figs. 2(a) and 2(b) show that if an elastic model is to work for metallic glasses with a universal value of the characteristic volume $V_c$ in terms of the molar volume $V_m$, the bulk modulus can play only a minor role. This constitutes an experimental demonstration of ``shear dominance''. 

For the energy-landscape version of the shoving model Eq. (\ref{de}) translates into the prediction

\begin{equation}\label{shv2}
T_g\propto 
\Delta E(T_g)\propto GV_m\frac{K+4G/3}{2K+11G/3}\, \,.
\end{equation}
This is compared to experiment in Fig. 2(c). The fit is good and there is no correlation to the Poisson ratio.

The three ``zero-parameter'' elastic models of Fig. 2 have standard deviations, respectively, of 8\% for the standard shoving model [Fig. 2(a)], 22\%  for the bulk modulus elastic model [Fig. 2(b)], and 8\% for the energy-landscape version of the shoving model [Fig. 2(c)]. The latter gives no better fit to the data than the original shoving model (the standard deviation of it is 0.6\% lower, but this is hardly significant), but has the advantage of eliminating the correlation to the Poisson ratio.

A pragmatic one-parameter version of the elastic models is to allow for an arbitrary combination of $G$ and $K$ by assuming that $T_g$ is controlled by $\alpha G + (1-\alpha) K$. Due to shear dominance one expects $\alpha$ to be close to unity. Indeed, Ref. \onlinecite{wan12} showed that for $\alpha=10/11$ a very good fit to data is obtained for metallic glasses, a fit which like the energy-landscape shoving model eliminates the correlation to the Poisson ratio. For this one-parameter model the standard deviation is 6\%, i.e., somewhat better than the standard shoving model and its energy-landscape version.

\begin{figure}
  \centering
  \includegraphics{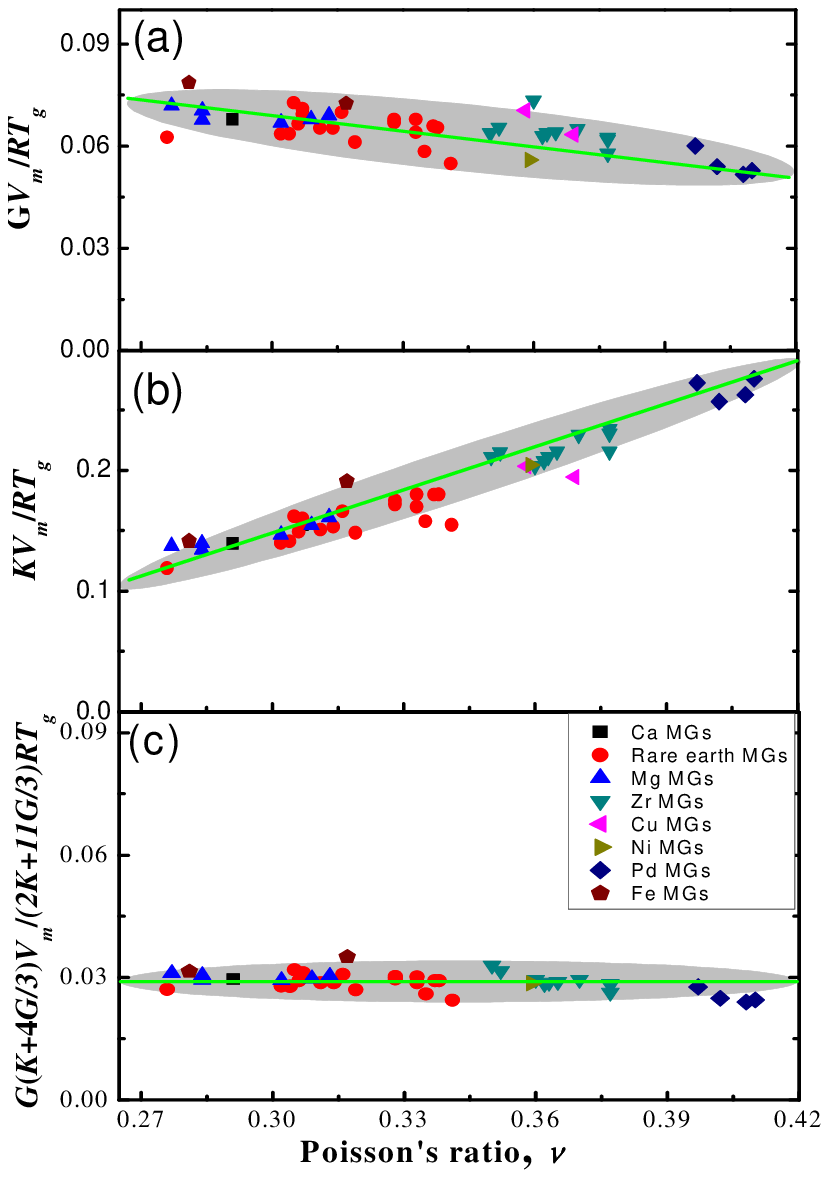}
   \caption{Comparing shoving model predictions to data for a range of metallic glasses (Fig. 1). In order to reduce the number of free parameters as much as possible it is assumed that the characteristic volumes $V_c$ and $V_c'$ via Eqs. (\ref{shov}) and (\ref{de}) are proportional to the molar volume $V_m$ with the same constant of proportionality for all compositions. 
(a) tests the original shoving model (Eq. (\ref{shov})), according to which the quantity $GV_m/T_g$ is the same for all systems, where $G$ is the glass shear modulus. This is obeyed to a fairly good approximation, but there is a slight correlation with the glass' Poisson ratio.\cite{wan12} This correlation is opposite and considerably stronger for the bulk modulus controlled elastic model [(b)]. The energy-landscape analogue of the shoving model [Eq. (\ref{de})], which is tested in (c), does not show correlation with the Poisson ratio.}
\end{figure}

\section{Summary}

The instantaneous shear modulus of the shoving model refers to the plateau modulus that is traditionally in experiment denoted by $G_\infty$. This quantity is measurable, in contrast to the idealized affine instantaneous shear modulus of liquid state theory, which was also traditionally denoted by $G_\infty$. A consistent notation has been suggested for distinguishing between these two quantities whenever there is a risk of confusing them. Data for a range of metallic glasses have been shown to be consistent with shoving model predictions; in particular the energy-landscape version of the model eliminates the correlation of model predictions to Poisson's ratio.

\acknowledgments 
The centre for viscous liquid dynamics ``Glass and Time'' is sponsored by the Danish National Research Foundation (DNRF).

\end{document}